\documentclass{ifacmtg}
\usepackage[dcucite]{harvard}
\usepackage[dvips]{epsfig}
\usepackage{rotate}
\begin{document}
\runauthor{D. Helbing}
\runtitle{Traffic data and consistent models}
\begin{frontmatter}
\title{Traffic Data and Their Implications for 
Consistent Traffic Flow Modeling}
\author[Stuttgart]{Dirk Helbing\thanksref{Someone}}
\address[Stuttgart]{II. Institute of Theoretical Physics,
University of Stuttgart, Pfaffenwaldring 57/III, 70550 Stuttgart, Germany,
www:~http://www.uni-stuttgart.de/UNIuser/thphys/helbing.html}
\thanks[Someone]{The author is grateful to Henk Taale and the
Ministry of Transport, Public Works and Water Management for supplying the
freeway data.}
\begin{abstract}
The paper analyzes traffic data 
with respect to
certain aspects which are relevant for traffic flow modeling as well as
the calibration of model parameters and functions. Apart from the
dynamic velocity distribution, the density-dependence and the temporal
evolution of various, partly lane-specific quantities is investigated.
The results are well compatible with recent macroscopic traffic flow models which 
have been derived from the dynamics of driver-vehicle units. These have also
solved the inconsistencies, which previous models have been criticized for. 
\end{abstract}
\begin{keyword}
Calibration, consistency, evaluation, mathematical models,
vehicle dynamics
\end{keyword}
\end{frontmatter}

\section{Introduction}

Efficient traffic flow is very important for a high standard of living.
However, during the last decades the traffic volume has continuously
increased, so that the capacity of the road system is almost reached in many
countries, especially in conurbations. Therefore, methods for
traffic flow optimization are called for. To develop these, numerous
traffic simulation models have been proposed. 
\par
This paper will focus on
macroscopic (fluid-dynamic) models for the vehicle density and the
average velocity. It addresses essentially two questions: 1. Which
phenomena and functional relations should be reproduced by the models?
Related results are presented in the next section. 2. Which one of the existing
traffic flow models should be used to simulate traffic dynamics?
This is the topic of Sect.~\ref{S2}. In particular, it is discussed which
consistency requirements should be met by realistic models.

\section{Data evaluation} \label{S1}
During the last year single vehicle data of the Dutch two-lane
Autobahn A9 from Haarlem to Amsterdam 
have been evaluated with
respect to various criteria which are relevant for traffic flow
modeling. The quantities measured by induction loops at discrete places $x$
along the freeway
include the passage times $t_\alpha(x)$, velocities $v_\alpha(x)$, vehicle lengths
$l_\alpha(x)$, and lanes $i_\alpha(x)$ of the single vehicles $\alpha$.
These allow the determination of various aggregate (so-called macroscopic)
quantities. 
\par
The results of the standard evaluation methods have already been
presented, compared and discussed elsewhere \cite{Hel97}. They base on averages over
a fixed time interval $T$, giving large statistical variations 
of the velocity moments
\begin{equation}
 \langle v^k \rangle = \frac{1}{N(x,t)} \sum_{t-T/2 \le t_\alpha < t + T/2}
 [v_\alpha(x)]^k
\end{equation}
where the traffic flow ${\cal Q}(x,t) =
N(x,t)/T$ is small, since the number $N(x,t)$ of vehicles
passing the intersection at place $x$ between times $t-T/2$ and $t + T/2$ is
small, then. 
\par
The following considerations will focus on an evaluation method with a sampling
error that is independent of the traffic volume. For this purpose, one 
averages over a fixed number $N=100$ of vehicles, implying a variation
of the time interval $T(x,t)$ of data collection. Then, the definition of the
space- and time-dependent traffic flow is ${\cal Q}(x,t) = N/T(x,t)$ with
\begin{equation}
 t = \frac{1}{N} \sum_{\alpha_0\le \alpha <\alpha_0+N} t_\alpha(x) \, ,
\end{equation}
and the velocity moments are given by
\begin{equation}
 \langle v^k \rangle = \frac{1}{N} \sum_{\alpha_0\le \alpha <\alpha_0+N}
 [v_\alpha(x)]^k \, . 
\end{equation}
This allows the calculation of the average velocity $V(x,t) = \langle v
\rangle$, the density $\rho(x,t) = {\cal Q}(x,t) / V(x,t)$, the velocity variance
\begin{equation}
 \Theta(x,t) = \langle [v - V(x,t)]^2 \rangle = \langle v^2 \rangle
 - \langle v \rangle^2 \, ,
\end{equation}
the skewness
\begin{equation}
 \gamma(x,t) = \frac{\langle [v - V(x,t)]^3 \rangle}{[\Theta(x,t)]^{3/2}} \, ,
\end{equation}
and the kurtosis
\begin{equation}
 \delta(x,t) 
 = \frac{\langle [v - V(x,t)]^4 \rangle}{[\Theta(x,t)]^{2}} -3 \, .
\end{equation}
The definitions for the single lanes $i$ are analogous to the above ones
for the total cross section.
\par
\begin{figure}[htbp]
\unitlength5mm
\begin{center}
\begin{picture}(16,10.4)(0.6,-0.8)
\put(0,9.8){\epsfig{height=16\unitlength, width=9.8\unitlength, angle=-90, 
      bbllx=50pt, bblly=50pt, bburx=554pt, bbury=770pt, 
      file=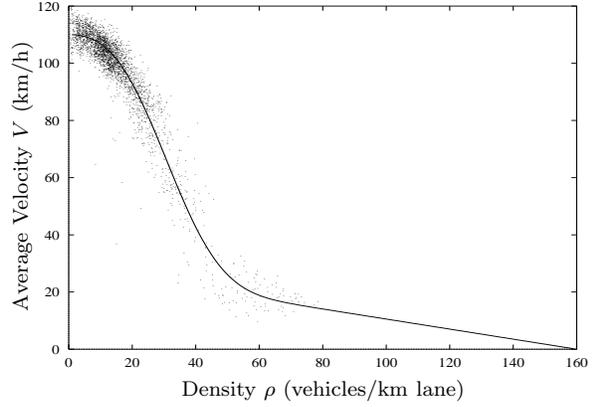}}
\put(8.7,-0.6){\makebox(0,0){\footnotesize Density $\rho$ (vehicles/km lane)}}
\put(0.7,5.1){\makebox(0,0){\rotate[l]{\hbox{\footnotesize Average Velocity 
$V$ (km/h)}}}}
\end{picture}
\end{center}
\caption[]{Empirical velocity-density data ($\cdot$) for the Dutch
  two-lane freeway A9 with a speed limit of 120\,km/h. The fit function 
$V_e(\rho)$ \mbox{(---)}
at high densities is reconstructed by means of theoretical traffic
flow relations (cf. Sect.~\ref{S2}).\label{f1}}
\end{figure}
\begin{figure}[htbp]
\unitlength5mm
\begin{center}
\begin{picture}(16,10.4)(0.6,-0.8)
\put(0,9.8){\epsfig{height=16\unitlength, width=9.8\unitlength, angle=-90, 
      bbllx=50pt, bblly=50pt, bburx=554pt, bbury=770pt, 
      file=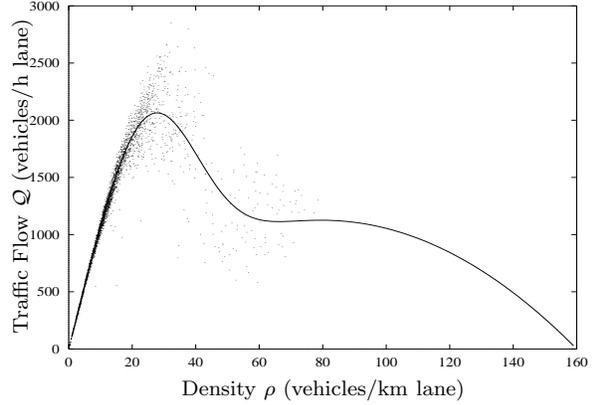}}
\put(8.7,-0.6){\makebox(0,0){\footnotesize Density $\rho$ (vehicles/km lane)}}
\put(0.7,5.1){\makebox(0,0){\rotate[l]{\hbox{\footnotesize Traffic Flow ${\cal Q}$ 
(vehicles/h lane)}}}}
\end{picture}
\end{center}
\caption[]{Empirical flow-density data ($\cdot$) and fit of the
fundamental diagram ${\cal Q}_e(\rho)$ (---).\label{f2}}
\end{figure}

Whereas the variance $\Theta$ is a measure for the breadth of the
velocity distribution $P(v;x,t)$, the skewness is a measure for its
asymmetry, and the kurtosis for its flatness. 
Most previous studies 
have restricted to the evaluation of the
velocity-density relation $V_e(\rho)$ (Fig.~\ref{f1})
and the fundamental diagram ${\cal Q}_e(\rho)$ (Fig.~\ref{f2}).
\par
\begin{figure}[htbp]
\unitlength4.9mm
\begin{center}
\begin{picture}(16,10.6)(0,-0.8)
\put(-0.4,9.8){\epsfig{height=16\unitlength, width=9.8\unitlength, angle=-90, 
      bbllx=50pt, bblly=50pt, bburx=554pt, bbury=770pt, 
      file=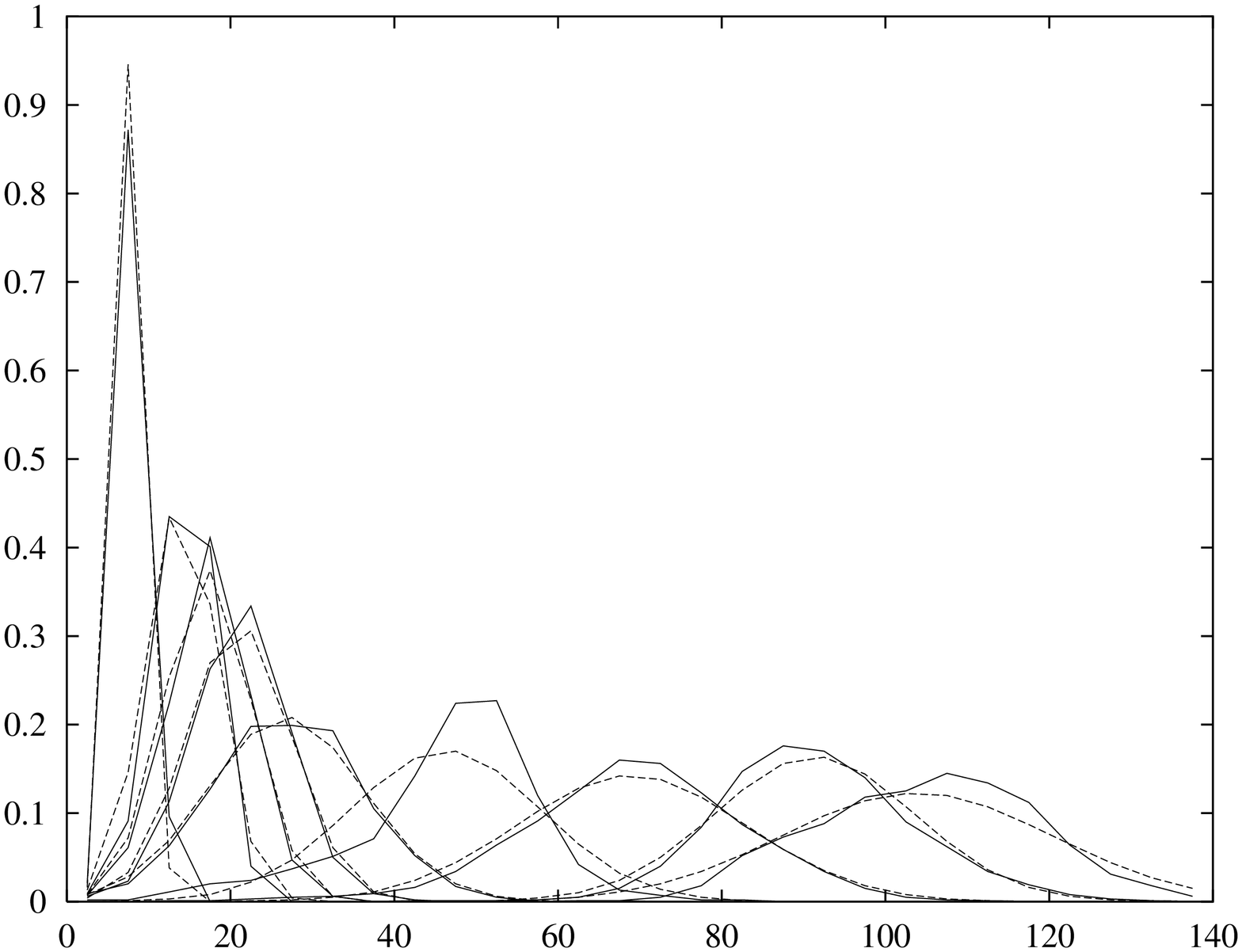}}
\put(8.7,-0.6){\makebox(0,0){\footnotesize Velocity $v$ (km/h)}}
\put(0.2,5.1){\makebox(0,0){\rotate[l]{\hbox{\footnotesize Average Velocity
        Distribution $P(v)$}}}}
\put(5.5,8){\makebox(0,0){\tiny $\rho=90$\,vehicles/km lane}}
\put(3.43,4.7){\makebox(0,0){\tiny 80}}
\put(4.24,4){\makebox(0,0){\tiny 70}}
\put(4.69,3.2){\makebox(0,0){\tiny 60}}
\put(5.15,2.5){\makebox(0,0){\tiny 50}}
\put(7.44,2.5){\makebox(0,0){\tiny 40}}
\put(8.56,2.5){\makebox(0,0){\tiny 30}}
\put(10.44,2.5){\makebox(0,0){\tiny 20}}
\put(12.15,2.2){\makebox(0,0){\tiny 10}}
\end{picture}
\end{center}
\caption[]{Comparison of empirical velocity distributions at different
densities (---) with frequency polygons of 
grouped normal distributions having
the same mean value and variance (--~--). 
A significant deviation of the empirical relations from the
respective discrete normal approximations is only found at 
$\rho = 40$\,vehicles/km lane, where the averaging interval 
$T$ may have been too long due to rapid stop-and-go waves
(cf.\ the mysterious ``knee'' at $\rho \approx 40$\,vehicles/km lane 
in Fig.~\ref{f7}).\label{f3}}
\end{figure}

\begin{figure}[htbp]
\unitlength4.8mm
\begin{center}
\begin{picture}(16,11.5)(0.6,-0.8)
\put(0.4,-0.6){\epsfig{width=16\unitlength, height=12\unitlength, angle=0, 
      file=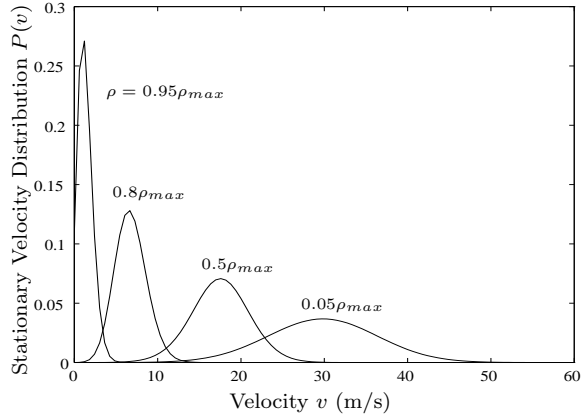}}
\put(8.7,-0.6){\makebox(0,0){\footnotesize Velocity $v$ (m/s)}}
\put(0.7,5.3){\makebox(0,0){\rotate[l]{\hbox{\footnotesize Stationary Velocity
        Distribution $P(v)$}}}}
\put(4.7,8){\makebox(0,0){\tiny $\rho=0.95\rho_{max}$}}
\put(4.24,5.2){\makebox(0,0){\tiny $0.8\rho_{max}$}}
\put(6.7,3.2){\makebox(0,0){\tiny $0.5\rho_{max}$}}
\put(9.6,2.1){\makebox(0,0){\tiny $0.05\rho_{max}$}}
\end{picture}
\end{center}
\caption[]{Stationary velocity distributions at different vehicle densities 
obtained from a improved kinetic traffic model (reproduction by kind
permission of C. Wagner).\label{f4}}
\end{figure}

However, for some theoretical considerations, it is essential to know the
form of the velocity distribution in dependence of density. In agreement with
previous studies 
it was found that a normal distribution is
a good approximation for the average velocity distribution 
(Fig.~\ref{f3}). Recent numerical solutions for the stationary velocity
distribution of an improved kinetic traffic model yield similar results
(Fig.~\ref{f4}). 
\par
\begin{figure}[htbp]
\unitlength5mm
\begin{center}
\begin{picture}(16,10.4)(0.6,-0.8)
\put(-0.2,9.8){\epsfig{height=16\unitlength, width=9.8\unitlength, angle=-90, 
      bbllx=50pt, bblly=50pt, bburx=554pt, bbury=770pt, 
      file=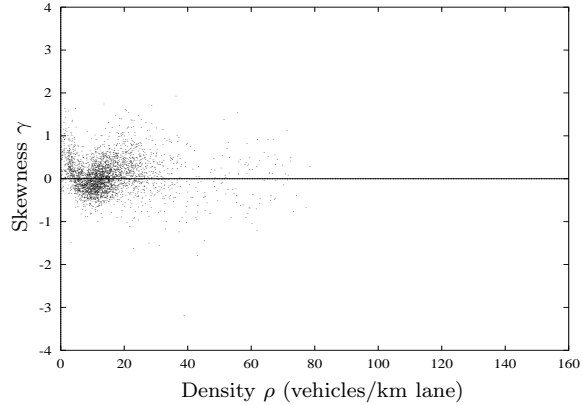}}
\put(8.7,-0.6){\makebox(0,0){\footnotesize Density $\rho$ (vehicles/km lane)}}
\put(0.7,5.1){\makebox(0,0){\rotate[l]{\hbox{\footnotesize Skewness
$\gamma$}}}}
\end{picture}
\end{center}
\caption[]{The empirical skewness ($\cdot$) is approximately zero most of the
time.\label{f5}}
\end{figure}

\begin{figure}[htbp]
\unitlength5mm
\begin{center}
\begin{picture}(16,10.4)(0.6,-0.8)
\put(-0.2,9.8){\epsfig{height=16\unitlength, width=9.8\unitlength, angle=-90, 
      bbllx=50pt, bblly=50pt, bburx=554pt, bbury=770pt, 
      file=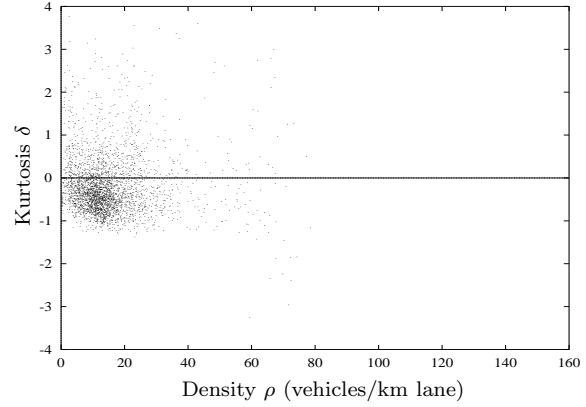}}
\put(8.7,-0.6){\makebox(0,0){\footnotesize Density $\rho$ (vehicles/km lane)}}
\put(0.7,5.1){\makebox(0,0){\rotate[l]{\hbox{\footnotesize Kurtosis $\delta$}}}}
\end{picture}
\end{center}
\caption[]{The empirical kurtosis shows a small systematic
  deviation from zero due the finite truck fraction. Nevertheless, it is
almost zero.\label{f6}}
\end{figure}

Due to the limited amount of data,
the {\em time-dependent} velocity distribution is hard to
obtain. Therefore, it has been investigated whether the time-dependent 
skewness (Fig.~\ref{f5}) and the kurtosis (Fig.~\ref{f6})
vanish in agreement with a normal distribution. The deviation of the
skewness from zero is indeed neither systematic nor significant. The kurtosis
is also very small but, on average, somewhat smaller than zero. Since this
systematic deviation from zero 
can be shown to vanish on the left-hand lane, it is probably an
effect of the finite truck fraction driving slower than other vehicles.
\par
In summary, the velocities are almost normally distributed, even in the
dynamic case. Nevertheless, if one averages over too large
time intervals $T$, bimodal distributions may result in cases of
strong variations of traffic flow. 
\par
\begin{figure}[htbp]
\unitlength5mm
\begin{center}
\begin{picture}(16,10.4)(0.6,-0.8)
\put(-0.2,9.8){\epsfig{height=16\unitlength, width=9.8\unitlength, angle=-90, 
      bbllx=50pt, bblly=50pt, bburx=554pt, bbury=770pt, 
      file=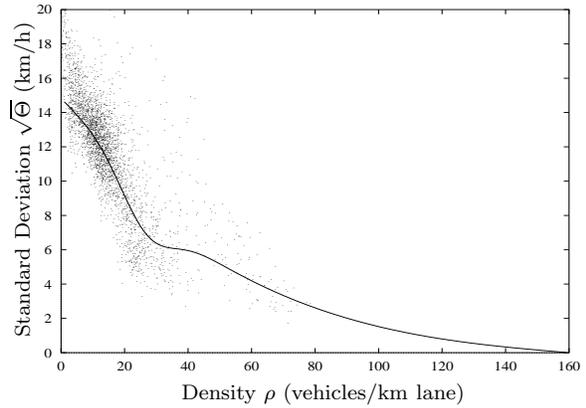}}
\put(8.7,-0.6){\makebox(0,0){\footnotesize Density $\rho$ (vehicles/km lane)}}
\put(0.7,5.1){\makebox(0,0){\rotate[l]{\hbox{\footnotesize Standard Deviation 
$\sqrt{\Theta}$ (km/h)}}}}
\end{picture}
\end{center}
\caption[]{Standard deviation $\sqrt{\Theta(x,t)}$ of vehicle velocities
in dependence of density $\rho(x,t)$ ($\cdot$) and corresponding fit function 
$\sqrt{\Theta_e(\rho)}$ (---).\label{f7}}
\end{figure}

\begin{figure}[htbp]
\unitlength5.4mm
\begin{center}
\begin{picture}(15.2,10.4)(1.4,-0.8)
\put(-0.2,9.8){\epsfig{height=16\unitlength, width=9.8\unitlength, angle=-90, 
      bbllx=50pt, bblly=50pt, bburx=554pt, bbury=770pt, 
      file=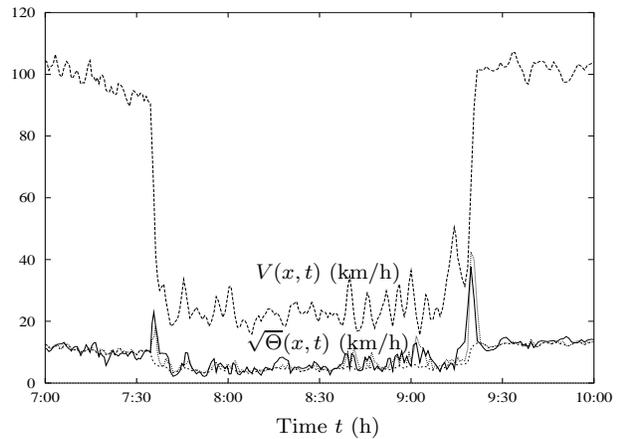}}
\put(8.7,-0.6){\makebox(0,0){\footnotesize Time $t$ (h)}}
\put(8.7,3.2){\makebox(0,0){\footnotesize $V(x,t)$ (km/h)}}
\put(8.7,1.5){\makebox(0,0){\footnotesize $\sqrt{\Theta}(x,t)$ (km/h)}}
\end{picture}
\end{center}
\caption[]{The standard deviation $\sqrt{\Theta(x,t)}$ of vehicle velocities
(---) has maxima, when the average velocity $V(x,t)$ (--~--) changes
considerably. It cannot be described by the variance-density relation
$\sqrt{\Theta_e(\rho(x,t))}$ alone (-~-~-), but together with an
additional positive term which arises from the evaluation procedure
($\cdots$).\label{f8}} 
\end{figure}

Moreover, it is interesting to know the variance-density relation
(Fig.~\ref{f7}). Looking at the {\em temporal} evolution of the
variance, one detects significant peaks when the average velocity 
breaks down or recovers (Fig.~\ref{f8}). 
Detailled mathematical considerations show \cite{Hel97}, that it can be
reconstructed from the temporal course of the vehicle density $\rho(x,t)$ via the
variance-density relation $\Theta_e(\rho)$, 
if one takes into account a positive correction term
which is caused by averaging over finite time periods $T$. The theoretical 
relation reads
\begin{equation}
 \Theta(x,t) \approx \Theta_e(\rho(x,t)) + \frac{T^2}{4}
 \left( \frac{\partial V}{\partial t} \right)^2 \, .
\label{Eq7}
\end{equation}
\par
\par
\begin{figure}[htbp]
\unitlength5mm
\begin{center}
\begin{picture}(16,10.4)(0.6,-0.8)
\put(-0.2,9.8){\epsfig{height=16\unitlength, width=9.8\unitlength, angle=-90, 
      bbllx=50pt, bblly=50pt, bburx=554pt, bbury=770pt, 
      file=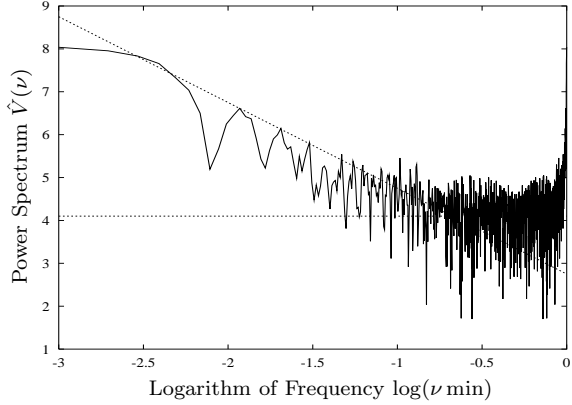}}
\put(8.7,-0.6){\makebox(0,0){\footnotesize Logarithm of Frequency
$\log (\nu$\,min)}}
\put(0.7,5.1){\makebox(0,0){\rotate[l]{\hbox{\footnotesize Power Spectrum
$\hat{V}(\nu)$}}}}
\end{picture}
\end{center}
\caption[]{Power spectrum of the average velocity's temporal course
(-~-~-: $\log C - \delta \log \nu$, $\delta \in \{0,2\}$).\label{f10}}
\end{figure}

Investigating the temporal fluctuations of the 
average velocity $V(x,t)$, one finds a flat power spectrum
$\hat{V}(\nu)$, i.e. almost the same 
intensity of the different fluctuation frequencies, corresponding to a so-called 
``white noise'' (Fig.~\ref{f10}).
However, at small frequencies $\nu$ (large oscillation periods $1/\nu$) 
the power law
\begin{equation}
 \hat{V}(\nu) = C \nu^{-2} 
\end{equation}
appears. Simular results are obtained for the fluctuations of the
density $\rho(x,t)$ 
or the variance $\Theta(x,t)$.
The power law behavior is most distinct near to 
entrances, but only weakly visible at undisturbed sections of the freeway. 
\par
\par
It has also been found that the dynamics on neighboring lanes is
strongly correlated\begin{figure}[htbp]
\unitlength5.4mm
\begin{center}
\begin{picture}(15.2,10.4)(1.4,-0.8)
\put(-0.2,9.8){\epsfig{height=16\unitlength, width=9.8\unitlength, angle=-90, 
      bbllx=50pt, bblly=50pt, bburx=554pt, bbury=770pt, 
      file=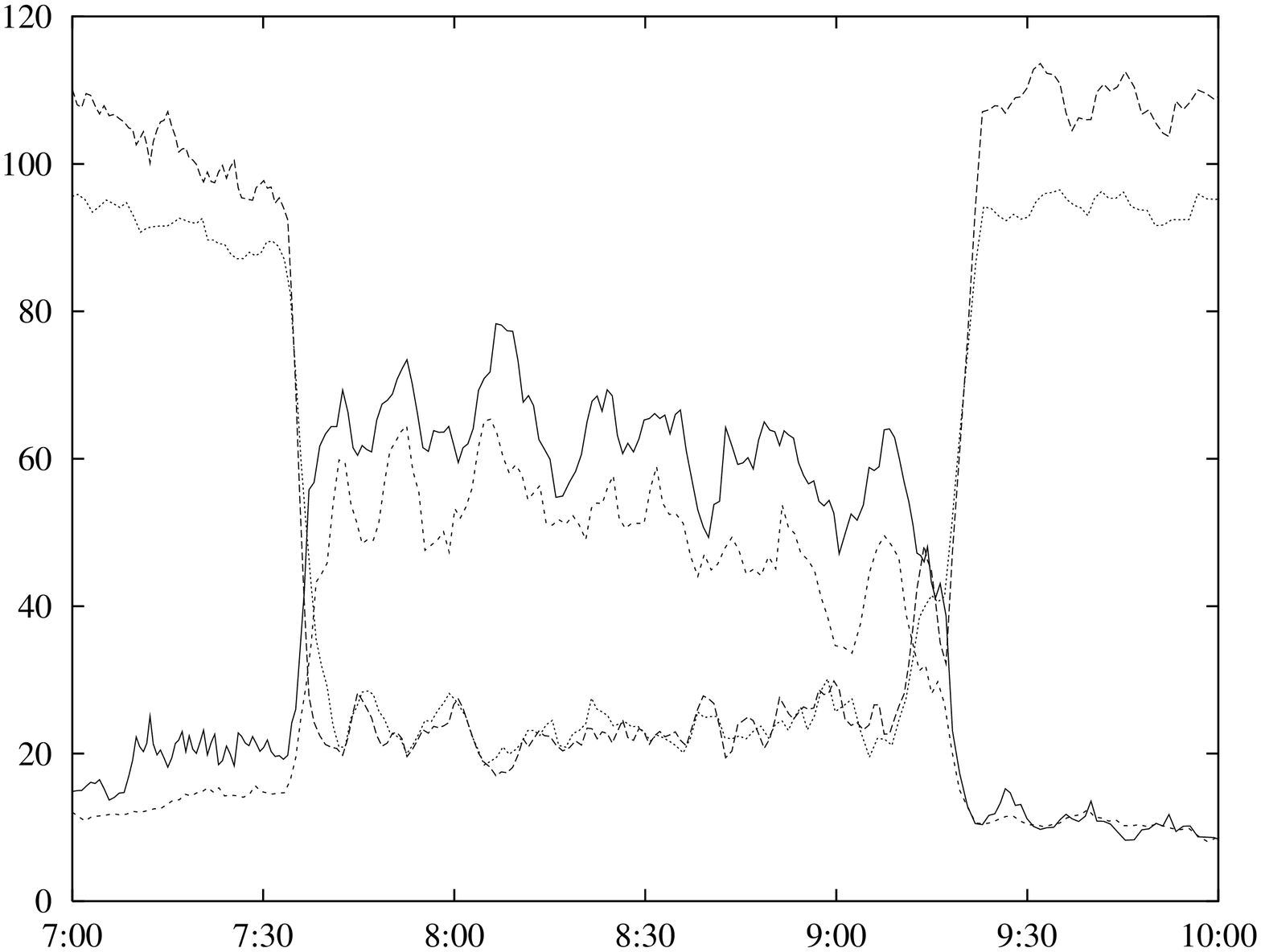}}
\put(8.7,-0.6){\makebox(0,0){\footnotesize Time $t$ (h)}}
\put(8.7,6.8){\makebox(0,0){\footnotesize $\rho_i(x,t)$ (vehicles/km lane)}}
\put(8.7,2.9){\makebox(0,0){\footnotesize $V_i(x,t)$ (km/h)}}
\end{picture}
\end{center}
\caption[]{Comparison of the temporal course of the vehicle density
$\rho_i(x,t)$ on the left lane \mbox{(---)} with the right lane (-~-~-) and
of the average velocity $V_i(x,t)$ on the left lane (--~--) with the right lane
($\cdots$).\label{f11}}
\end{figure}
(Fig.~\ref{f11}). This justifies the common practice
to model the dynamics of the total cross section of the road in an overall
manner, treating it like a single lane with higher capacity and possibilities
of overtaking (instead of explicitly describing the dynamics of {\em all} the
lanes, including their interactions by overtaking and lane-changing
maneuvers). Above an average density of 35 vehicles per kilometer 
and lane the velocity on the left lane is the same as on the right lane,
but the left lane is more crowded than the right one. 
The variance on neighboring lanes behaves almost identical \cite{Hel97}. 
\par
\begin{figure}[htbp]
\unitlength5mm
\begin{center}
\begin{picture}(16,10.4)(0.5,-0.8)
\put(0,9.8){\epsfig{height=16\unitlength, width=9.8\unitlength, angle=-90, 
      bbllx=50pt, bblly=50pt, bburx=554pt, bbury=770pt, 
      file=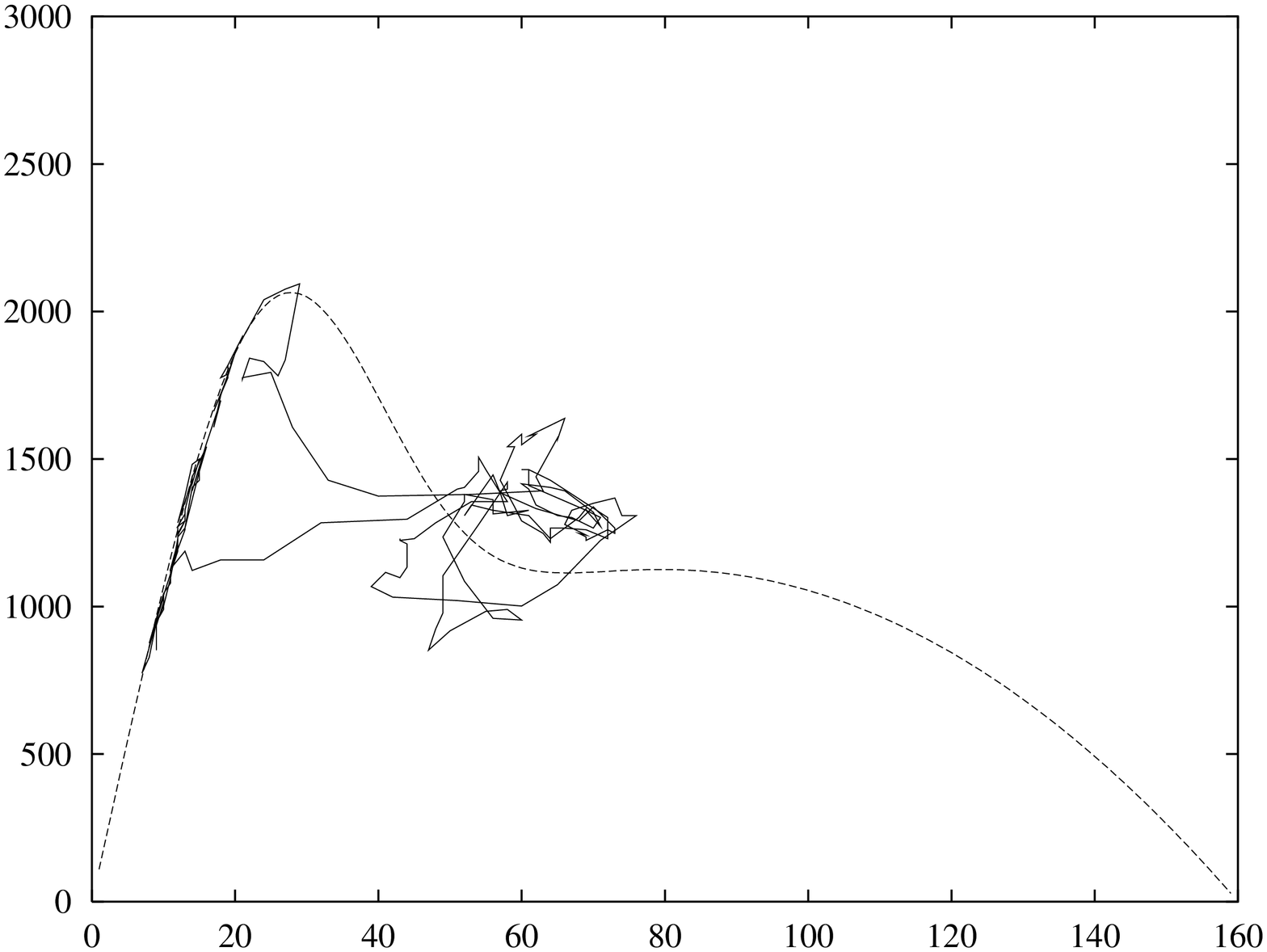}}
\put(8.7,-0.6){\makebox(0,0){\footnotesize Density $\rho$ (vehicles/km lane)}}
\put(0.7,5.1){\makebox(0,0){\rotate[l]{\hbox{\footnotesize Flow ${\cal Q}$
(vehicles/h)}}}}
\put(4.32,5.75){\vector(1,-2){0.15}}
\put(4,4.005){\vector(-1,0){0.2}}
\end{picture}
\end{center}
\caption[]{Comparison of the fundamental diagram ${\cal Q}_e(\rho)$
(--~--, cf. Fig.~\ref{f2}) with the temporal evolution of traffic flow
${\cal Q}(x,t)$ (---).\label{f12}}
\end{figure}

The maximum flow is 2200 vehicles per hour and lane, 
and it is reached at a density of
30 vehicles per kilometer and lane. Above this density, traffic flow
breaks down and stop-and-go traffic develops. However, the hysteresis curve
in Fig.~\ref{f12} indicates that traffic flow is already unstable
at densities of about 12 vehicles per kilometer and lane.
\par
\begin{figure}[htbp]
\unitlength5mm
\begin{center}
\begin{picture}(16,10.4)(0.6,-0.8)
\put(-0.2,9.8){\epsfig{height=16\unitlength, width=9.8\unitlength, angle=-90, 
      bbllx=50pt, bblly=50pt, bburx=554pt, bbury=770pt, 
      file=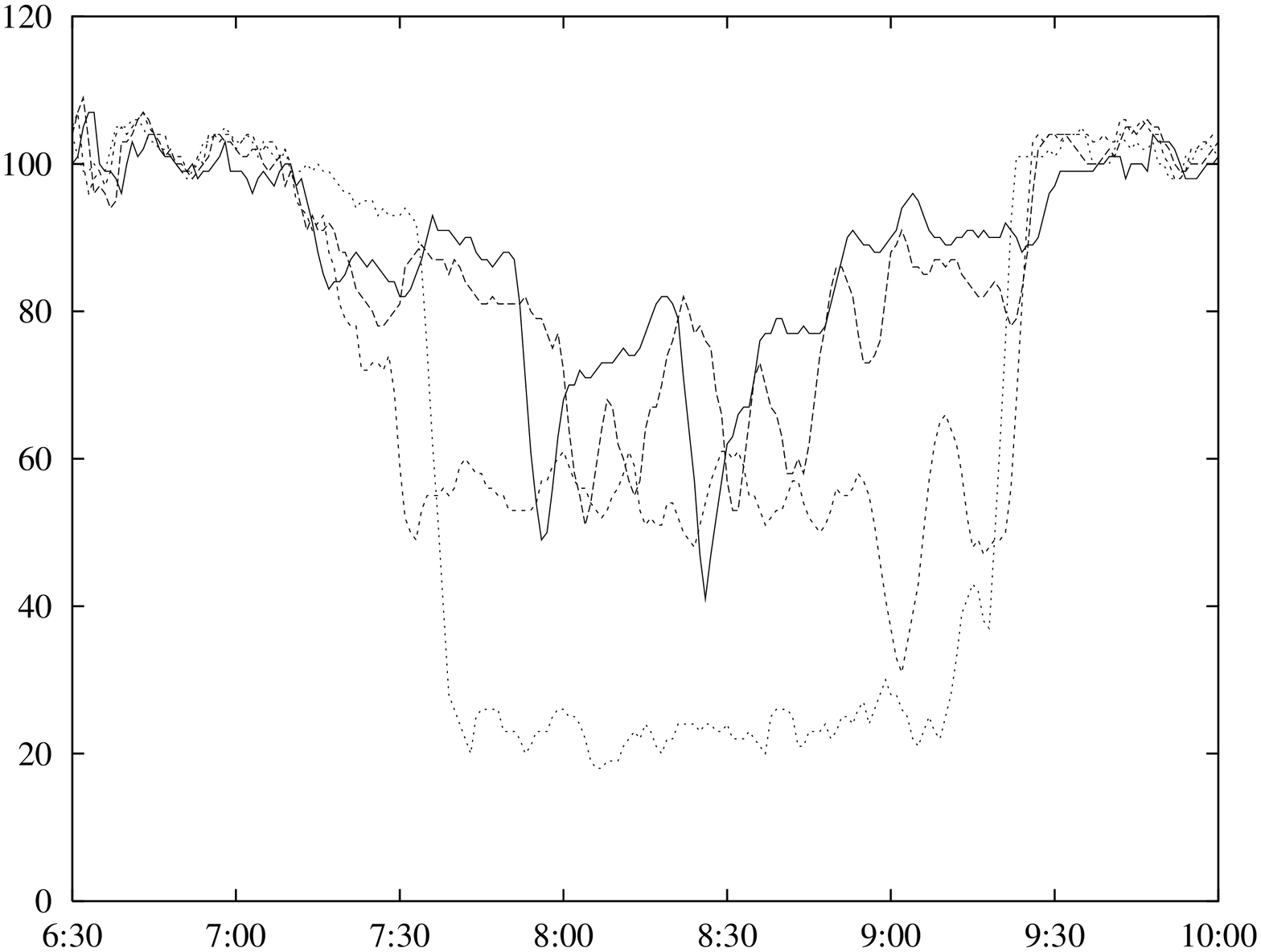}}
\put(8.7,-0.6){\makebox(0,0){\footnotesize Time $t$ (h)}}
\put(0.7,5.1){\makebox(0,0){\rotate[l]{\hbox{\footnotesize Average Velocity
$V(x,t)$ (km/h)}}}}
\end{picture}
\end{center}
\caption[]{Temporal evolution of the average velocity $V(x,t)$ at successive
  cross section of the freeway ($\cdots$: $x=0$\,km; -~-~-: $x=0.95$\,km;
--~--: $x=2.15$\,km; $x=4.15$\,km).\label{f13}}
\end{figure}

Although the average velocity $V(x,t)$ recovers after a bottleneck
(at $x=0$\,km), the amplitude of the 
emerging stop-and-go waves
becomes larger and larger (Fig.~\ref{f13}). 
At the same time their wave length grows, indicating a merging of traffic jams.
\par

\section{Consistent Traffic Modeling} \label{S2}

For a realistic description of traffic dynamics, it is sufficient to 
restrict the model to the continuity equation 
\begin{equation}
 \frac{\partial \rho}{\partial t}
 + \frac{\partial (\rho V)}{\partial x} = 0 
\end{equation}
for the vehicle density $\rho(x,t)$ and
a suitable dynamic equation for the average velocity $V(x,t)$. 
The velocity equation is necessary
for the delineation of emergent traffic jams and stop-and-go traffic.
The dynamics of the variance can be reconstructed from the temporal
evolution of the density and average velocity according to Eq.
(\ref{Eq7}). Due to the strong correlation
between neighboring lanes, an overall description of the total road
cross section is possible as long as the number of lanes does not change.
Nevertheless, models for the interaction between neighboring lanes
have also been built and simulated \cite{Hel97}.
\par
\begin{figure}[htbp]
\unitlength5mm
\begin{center}
\begin{picture}(16,10.4)(0.6,-0.8)
\put(-0.2,9.8){\epsfig{height=16\unitlength, width=9.8\unitlength, angle=-90, 
      bbllx=50pt, bblly=50pt, bburx=554pt, bbury=770pt, 
      file=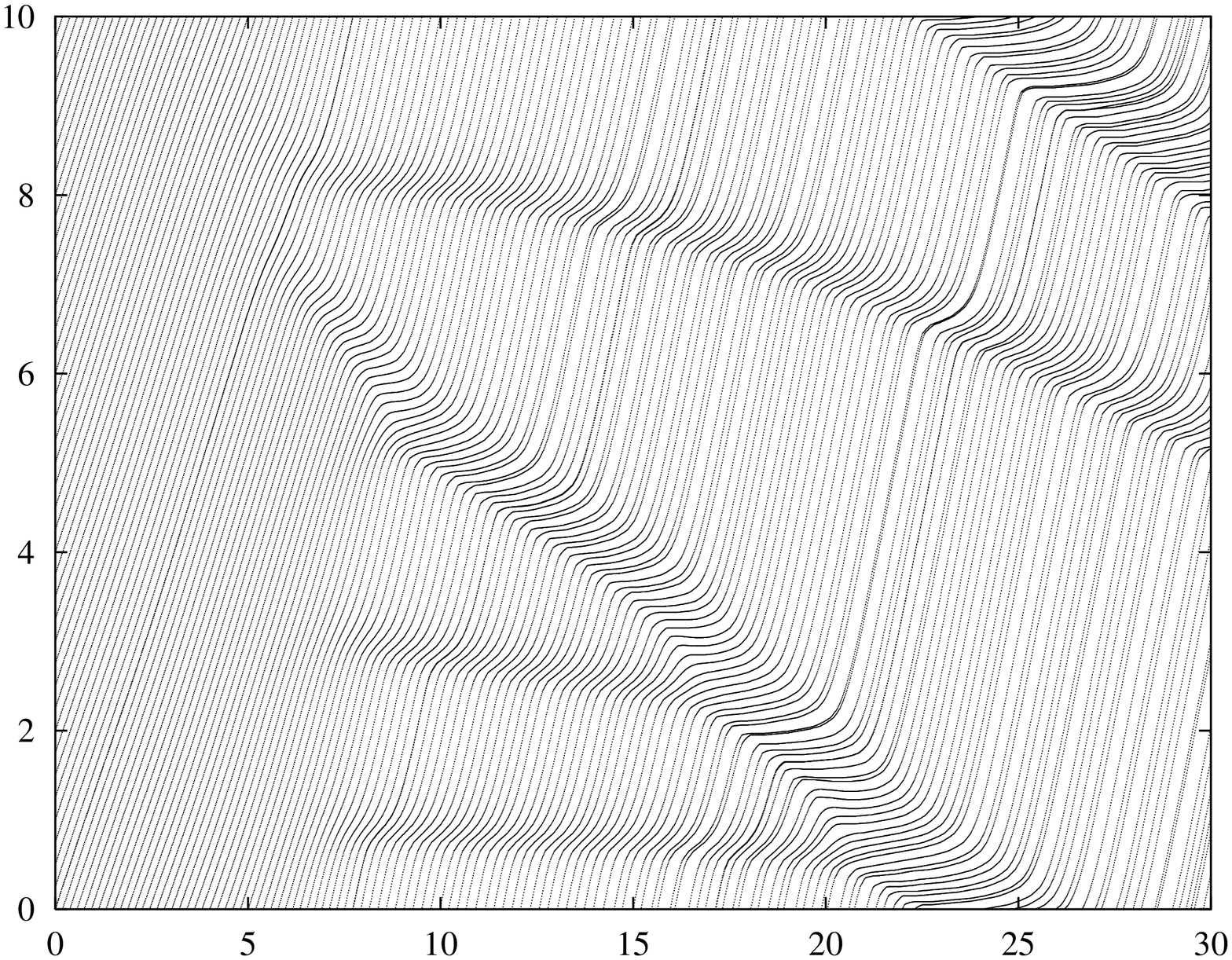}}
\put(8.7,-0.6){\makebox(0,0){\footnotesize Time $t$ (min)}}
\put(0.7,5.1){\makebox(0,0){\rotate[l]{\hbox{\footnotesize Place
$x$ (km)}}}}
\end{picture}
\end{center}
\caption[]{The representative streamlines related to the simulation of the
Kerner-Konh\"auser model indicate the average velocity by
their slope and the vehicle density by the line density. Starting with
an almost homogeneous traffic situation with a density of 40 vehicles
per kilometer and lane, density clusters (which correspond to traffic jams)
develop in the course of time. These merge into greater clusters
in agreement with Fig.~\ref{f13}.\label{f14}}
\end{figure}

It can be shown that the velocity equations of most previous macroscopic
traffic models (in their continuous version) can be written in the form
\begin{equation}
 \frac{\partial V}{\partial t} + V \frac{\partial V}{\partial x}
 = - \frac{1}{\rho} \frac{\partial {\cal P}}{\partial x}
 + \frac{1}{\tau} [ V_{e}(\rho) - V ] \, .
\label{vorher}
\end{equation} 
The main difference is the specification of the traffic pressure
${\cal P}$, the relaxation time $\tau$ and the velocity-density relation
$V_{e}(\rho)$. For example, the Lighthill-Whitham model
results in the limit $\tau \rightarrow 0$. Payne's and Papageorgiou's
model is obtained for ${\cal P} = - V_{e}/2\tau$. For
$d{\cal P}/d\rho = - \rho/[2\tau (\rho + \kappa)] dV_{e}/d\rho$
one ends up at Cremer's model. In the model of Phillips there is
${\cal P} = \rho \Theta$, where $\Theta$ denotes the velocity variance.
The model of K\"uhne, Kerner and Konh\"auser (Fig.~\ref{f14}) results for
${\cal P} = \rho \Theta_0 - \eta \partial V/\partial x$, where
$\Theta_0$ is a positive constant and $\eta$ a viscosity coefficient.
\par
All these models have proved their value in many applications, and it is
hard to decide which is the best one. Nevertheless, recently these models have
seriously been criticized by Daganzo and others. 
Therefore, a new macroscopic
model has been derived from Boltzmann-like kinetic traffic equations, basing
on a microscopic model of the dynamics of driver-vehicle units.
The kinetic traffic model is related to Paveri-Fontana's one, but 
it additionally takes into account vehicular space requirements (by applying 
the Enskog-formalism for dense gases)
and velocity fluctuations due to imperfect driving \cite{Hel97}. 
\par
The calculation of the corresponding macroscopic
traffic equations is very laborious. For the equilibrium velocity
one finds the relation
\begin{equation}
 V^e(\rho,V,\Theta) = V_0 - \frac{\tau(\rho) [1-p(\rho)] \rho \Theta}
 {1 - \rho/\rho_{max} - \rho T_r V} \, ,
\end{equation}
where $V_0$ is the average desired velocity,
$p(\rho)$ the probability of immediate overtaking, $\rho_{max}
\approx 160$ vehicles per kilometer and lane the maximum vehicle
density, and $T_r \approx 0.8$\,s the reaction time.
The equilibrium variance $\Theta^e(\rho,V,\Theta) = A(\rho) (V^2 + \Theta)$,
where $A(\rho)$ denotes the strength of velocity fluctuations,
vanishes with the average velocity $V$, as required. Utilizing these
relations, it is possible to reconstruct the 
velocity-density relation $V_e(\rho) = V^e(\rho,V^e,\Theta^e)$ (Fig.~\ref{f1})
and the variance-density relation $\Theta_e(\rho) = \Theta^e(\rho,V^e,\Theta^e)$
(Fig.~\ref{f7}) at high densities $\rho$ \cite{Hel97}. 
Applying the equilibrium approximation
\begin{eqnarray}
 \Theta(x,t) &\approx & \Theta^e(\rho(x,t),V(x,t),\Theta^e(x,t)) \nonumber \\
 &=& \frac{A(\rho(x,t)) [V(x,t)]^2}{1 - A(\rho(x,t))} \, ,
\end{eqnarray}
one finally arrives at the corrected velocity equation
\begin{eqnarray}
 \frac{\partial V}{\partial t} + V \frac{\partial V}{\partial x}
 &=& - \frac{1}{\rho} \frac{\partial {\cal P}}{\partial \rho}
 \frac{\partial \rho}{\partial x} + a \frac{\partial V}{\partial x}
- b \frac{\partial^2 \rho}{\partial x^2} \nonumber \\
 &+& \frac{\eta}{\rho}
 \frac{\partial^2 V}{\partial x^2} 
 + \frac{1}{\tau} [ V^{e}(\rho,V) - V ] .
\end{eqnarray} 
Comparing this with (\ref{vorher}), one obviously
obtains the additional terms $a\partial V/\partial x$
and $-b \partial^2 \rho/\partial x^2$. Moreover,
${\cal P}$, $\eta$, $V^{e}$, $a$, and $b$ are not only functions
of the density $\rho$ anymore, but they 
also depend on the average velocity $V$. The corresponding
functional relationships can be analytically derived, and the model
parameters have been determined from empirical data. 
\par
\begin{figure}[htbp]
\unitlength5mm
\begin{center}
\begin{picture}(16,10.4)(0.6,-0.8)
\put(-0.2,9.8){\epsfig{height=16\unitlength, width=9.8\unitlength, angle=-90, 
      bbllx=50pt, bblly=50pt, bburx=554pt, bbury=770pt, 
      file=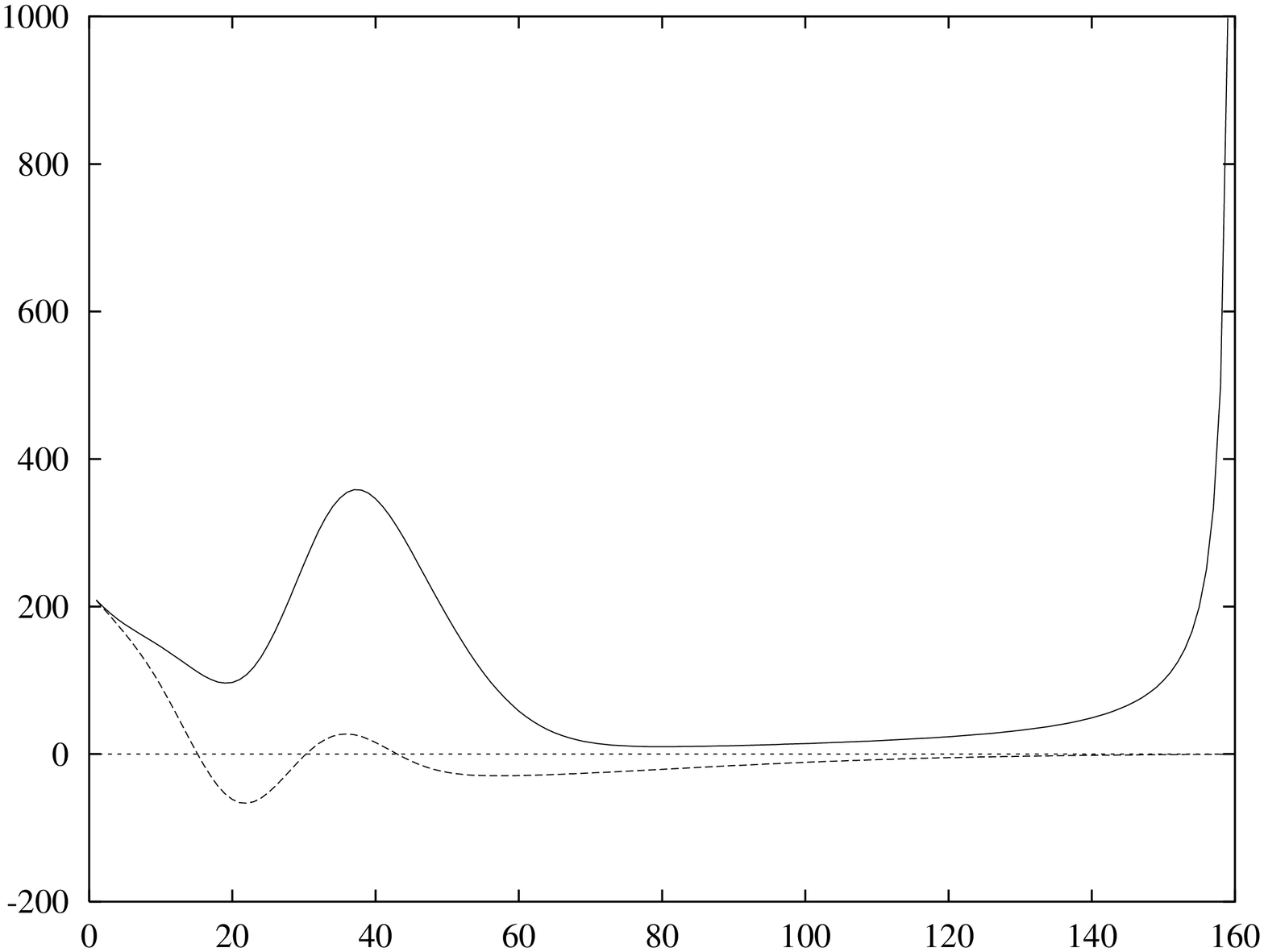}}
\put(8.7,-0.6){\makebox(0,0){\footnotesize Density $\rho$ (vehicles/km lane)}}
\put(0.7,5.1){\makebox(0,0){\rotate[l]{\hbox{\footnotesize
$\partial {\cal P}(\rho,V_e(\rho))/\partial \rho$ (km$^2$/h$^2$)}}}}
\end{picture}
\end{center}
\caption[]{Representation of the density-gradients of the
{\em idealized} traffic pressure ${\cal P} = \rho \Theta_{e}(\rho)$ 
of point-like vehicles (--~--) and the {\em corrected} pressure relation 
(---) which takes into account their finite space requirements.\label{f15}}
\end{figure}
\begin{figure}[htbp]
\unitlength5mm
\begin{center}
\begin{picture}(16,10.4)(0.6,-0.8)
\put(-0.2,9.8){\epsfig{height=16\unitlength, width=9.8\unitlength, angle=-90, 
      bbllx=50pt, bblly=50pt, bburx=554pt, bbury=770pt, 
      file=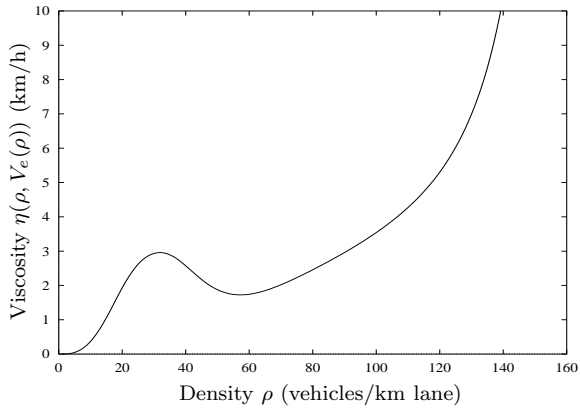}}
\put(8.7,-0.6){\makebox(0,0){\footnotesize Density $\rho$ (vehicles/km lane)}}
\put(0.7,5.1){\makebox(0,0){\rotate[l]{\hbox{\footnotesize Viscosity
$\eta(\rho,V_e(\rho))$ (km/h)}}}}
\end{picture}
\end{center}
\caption[]{Representation of the theoretically obtained viscosity 
function.\label{f16}}
\end{figure}
\begin{figure}[htbp]
\unitlength0.53cm
\begin{center}
\begin{picture}(14,6)(-0.4,0)
\put(0,6){\epsfig{height=14\unitlength, angle=-90, 
      bbllx=9cm, bblly=4.5cm, bburx=18.5cm, bbury=25cm, clip=,
      file=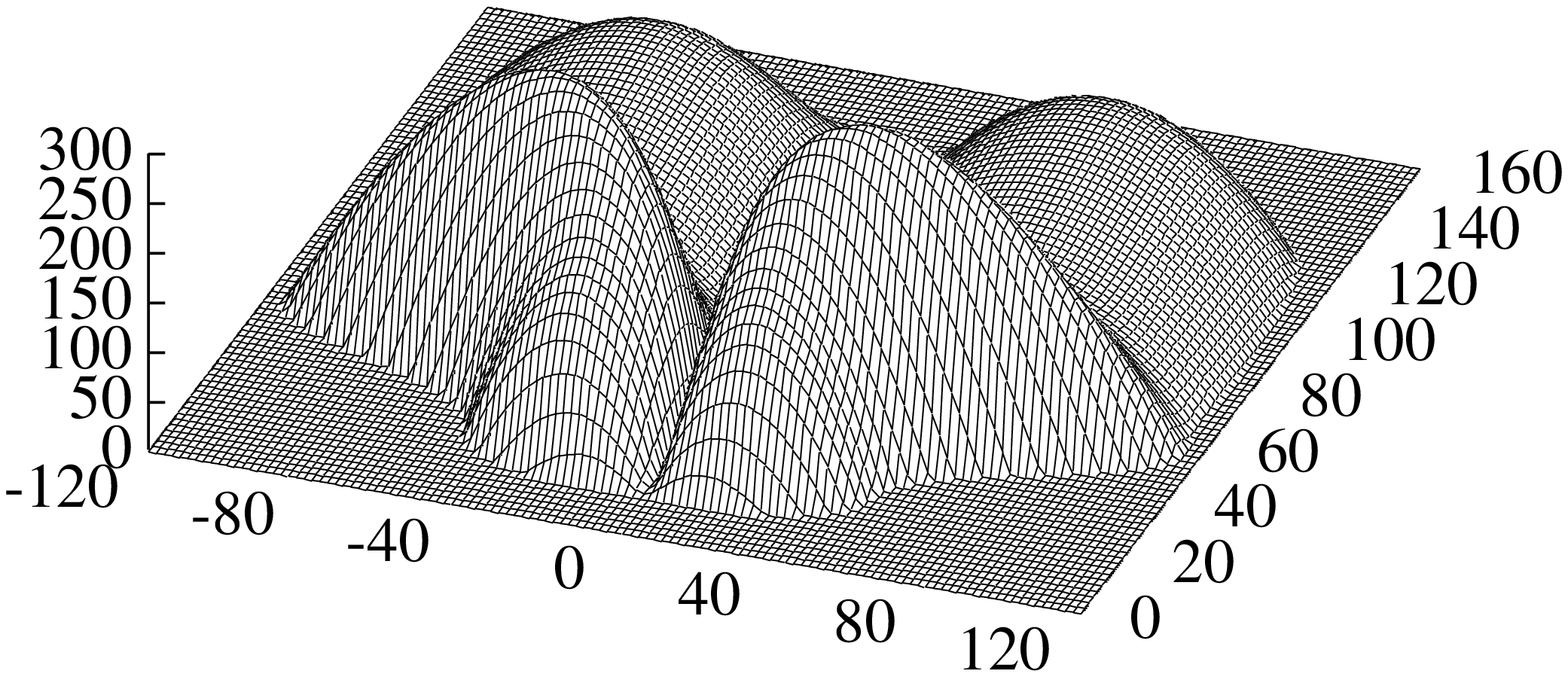}}
\put(1,6){\makebox(0,0){\footnotesize (a)}}
\put(1,5.2){\makebox(0,0){\footnotesize $\lambda(\rho,k)$ (1/h)}}
\put(4.8,-0.1){\makebox(0,0){\footnotesize $k$ (1/km)}}
\put(13.2,2.9){\makebox(0,0){\footnotesize $\rho$}}
\put(13,2.1){\makebox(0,0){\footnotesize (veh/}}
\put(12.8,1.3){\makebox(0,0){\footnotesize km lane)}}
\end{picture}
\begin{picture}(14,6.4)(-0.4,0)
\put(0,6.4){\epsfig{height=14\unitlength, angle=-90, 
      bbllx=8.4cm, bblly=4.5cm, bburx=18.5cm, bbury=25cm, clip=,
      file=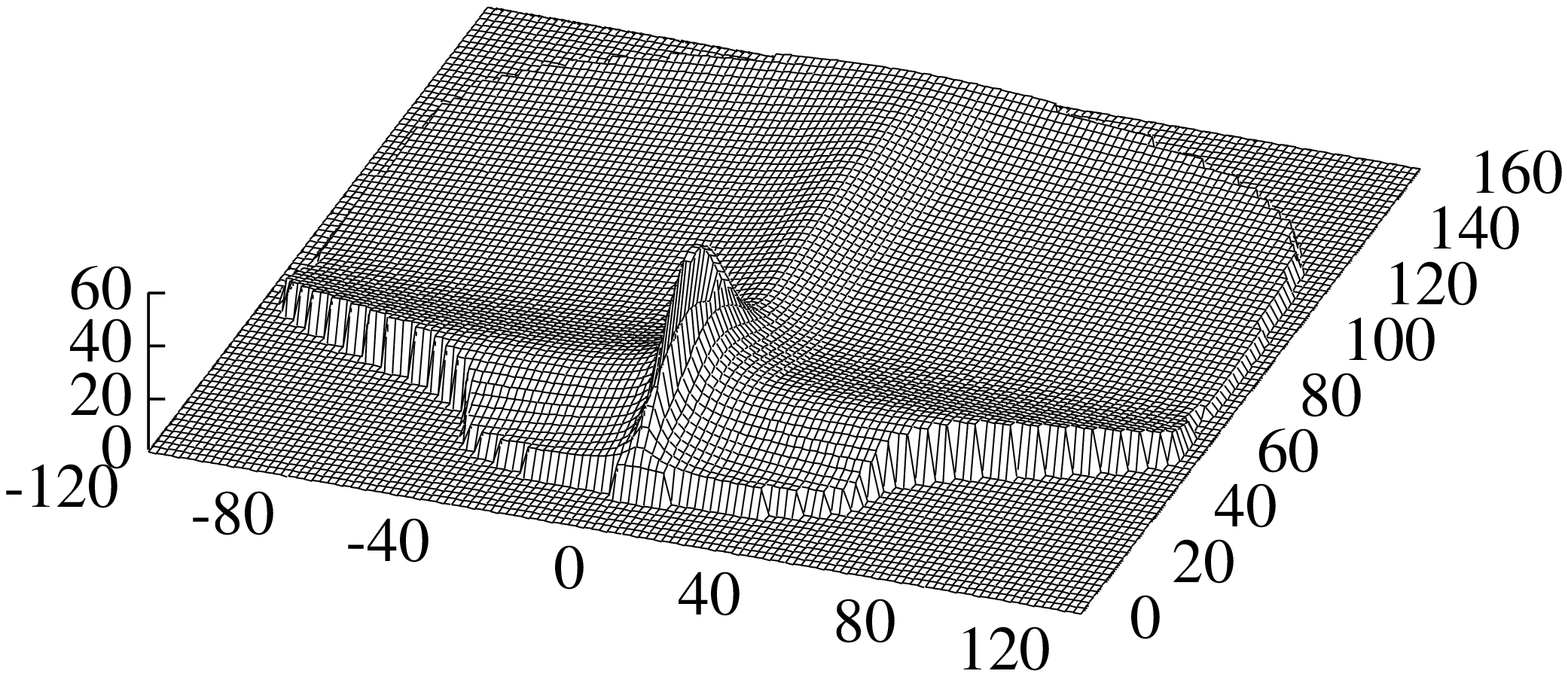}}
\put(1,5.6){\makebox(0,0){\footnotesize (b)}}
\put(1,4.6){\makebox(0,0){\footnotesize $c(\rho,k)$}}
\put(1,3.8){\makebox(0,0){\footnotesize (km/h)}}
\put(4.8,-0.1){\makebox(0,0){\footnotesize $k$ (1/km)}}
\put(13.2,2.9){\makebox(0,0){\footnotesize $\rho$}}
\put(13,2.1){\makebox(0,0){\footnotesize (veh/}}
\put(12.8,1.3){\makebox(0,0){\footnotesize km lane)}}
\end{picture}
\vskip1em
\end{center}
\caption[]{The illustrations show (a) the growth rate $\lambda$ 
and (b) the relative backward propagation
velocity $c$ of small periodic disturbances with wave number $k$
(corresponding to a wave length of $\ell = 2\pi/k$), where the 
homogeneous traffic flow at density $\rho$ is unstable.\label{f17}} 
\end{figure}

It can be shown that the model automatically meets all consistency criteria:
1.~The equilibrium velocity $V^{e}(\rho,V)$ decreases 
monotonically with growing density $\rho$ and vanishes at
the maximum vehicle density $\rho_{max}$.
2.~Since $\partial {\cal P}/\partial \rho$ and $b$ vanish for
$V=0$, the average velocity $V$ cannot become negative.
3.~ Since $\partial {\cal P}/\partial \rho$ is non-negative (Fig.~\ref{f15}),
the traffic pressure monotonically increases with the density $\rho$. As a
consequence, vehicles will not accelerate in direction of growing
density ($\partial \rho/\partial x > 0$).
4.~The traffic pressure diverges at $\rho = \rho_{max}$. For this
reason, the maximum density $\rho_{max}$ cannot be exceeded.
5.~The viscosity $\eta$ (Fig.~\ref{f16}) is theoretically explained. 
Its divergence at $\rho = \rho_{max}$ 
guarantees that velocity changes are smoothed out, so that the
development of shock-like structures is avoided and numerical integration
methods are stable.
6.~The traffic flow is found to be stable at small and extreme
densities, otherwise unstable (Fig.~\ref{f17}a). The result is in good
agreement with Fig.~\ref{f12}. Therefore, the development of
stop-and-go waves from almost homogeneous traffic conditions is
correctly described. The propagation velocity of the evolving density
waves (Fig.~\ref{f17}b) has the right order of magnitude and direction, i.e.
stop-and-go-waves move in backward direction with respect to the
average velocity $V$.

\section{Summary and conclusions}

Empirical traffic data have been compared to results of recent macroscopic
traffic models. The following has been found: 1. Vehicle velocities are
approximately normal distributed. This can be explained by an improved
kinetic model. 2. The dynamics on neighboring lanes is strongly correlated so
that the total freeway cross section can be described in an effective way.
3. The distance between forming stop-and-go waves increases in the course
of time, which corresponds to the merging of density clusters in traffic
simulations. 4. The dynamical variance has peaks when the velocity changes
considerably. It can be approximated by the variance-density relation and
an additional term which arises from averaging the vehicle data over finite
time intervals. 5. The fluctuations of traffic density and average velocity
essentially correspond to a white noise. 6. Traffic flow is unstable above
a density of about 12 vehicles per kilometer and lane. This can be
understood by means of a linear stability analysis of a new
macroscopic traffic flow model which has been derived from the dynamics
of driver-vehicle units via an improved kinetic model. 7. The model allows to
reconstruct the velocity-density and variance-density relations at high
densities and takes into account
velocity fluctuations as well as the finite space requirements of vehicles.
\bibliography{ifac}
\end{document}